\documentclass[
twocolumn,secnumarabic,amssymb, amsmath, tightenlines,nobibnotes, aps,longbibliography,superscriptaddress]
{revtex4-1}

\let\oldmarginpar\marginpar
\renewcommand\marginpar[1]{\-\oldmarginpar[\raggedleft\footnotesize #1]%
{\raggedright\footnotesize #1}}

\usepackage{color}			

\usepackage[normalem]{ulem}		

\usepackage{bm}		
\usepackage{amssymb,amsmath}
\usepackage{graphicx}

\usepackage{epstopdf}

\begin{document}

\title{Brownian duet:  A novel tale of thermodynamic efficiency}

\author{Karel Proesmans}
\email[email: ]{Karel.Proesmans@uhasselt.be}
\affiliation{Hasselt University, B-3590 Diepenbeek, Belgium}

\author{Yannik Dreher}
\author{Mom\v cilo Gavrilov}
\author{John Bechhoefer}
\affiliation{Department of Physics, Simon Fraser University, Burnaby, B.C., V5A 1S6, Canada}

\author{Christian Van den Broeck}
\affiliation{Hasselt University, B-3590 Diepenbeek, Belgium}

\begin{abstract}

We calculate analytically the stochastic thermodynamic properties of an isothermal Brownian engine driven by a duo of time-periodic forces, including its Onsager coefficients, the stochastic work of each force, and the corresponding stochastic entropy production. We verify the relations between different operational regimes, maximum power, maximum efficiency and minimum dissipation, and  reproduce the signature features of the stochastic efficiency. All these results are experimentally tested without adjustable parameters on a colloidal system.

\end{abstract}

\maketitle

\section{Introduction}

The efficiency of a machine is a central, founding principle in thermodynamics. Carnot realised that the impossibility of a perpetuum mobile (of the second kind) implies that the efficiency of a thermal machine is bounded from above and that this bound is universal, independent of the details in composition or construction of the engine \cite{carnot1824}. His key insight was that maximum efficiency is reached when operating in the immediate vicinity of an equilibrium point. A machine operating under this condition can function equally well as an engine and as a refrigerator. Carnot's work eventually led Clausius to introduce a new state function, the entropy, and to show that the upper bound of Carnot was tantamount to the famous  second law of thermodynamics, i.e., the increase of the total entropy \cite{clausius1867mechanical}.  

In the original Clausius formulation of thermodynamics, the entropy is well defined only for systems in {equilibrium,} and the second law refers to the increase of the total entropy of a closed system.  Through the work primarily of Ilya Prigogine, it became clear that one can also define entropy  in a  state of local equilibrium \cite{prigogine1967introduction,degroot11}. Together with Lars Onsager, he lay the groundwork for linear irreversible thermodynamics, a theory {that}  was further developed in detail for chemical and hydrodynamical systems \cite{glansdorff1973thermodynamic,degroot11,kondepudi2014modern}.  Onsager provided as {an} additional ingredient, the symmetry of a properly defined response, or Onsager matrix ${\bf L}$, ${\bf L}={\bf L}^T$, (for variables {that are} even under time-reversal).  {The symmetry derives} from the  reversibility of the microscopic laws.   As the off-diagonal elements of the Onsager matrix describe, for a thermal engine, the machine and refrigerator functions, respectively, this symmetry extends the {complementarity}, noted by Carnot, between these two modes of operations into the linear realm of irreversible thermodynamics.   Although numerous examples of Onsager symmetry have been documented  \cite{kondepudi2014modern}, they are mostly limited to {steady-state} systems.  Most thermodynamic machines, on the other hand---in particular the Carnot engine---operate under a time-periodic protocol.  Surprisingly, it is only recently  \cite{izumida2008molecular,izumida2009onsager,brandner2015thermodynamics,proesmans2015onsager,cerino2016linear,brandner2016periodic,bauer2016optimal,proesmans2015linear} that {such systems have been discussed} in the context of linear irreversible thermodynamics and that the corresponding Onsager matrix and its symmetry properties have been uncovered.  One significant novelty is that {the} Onsager symmetry is {generalized to the} Onsager-Casimir relation ${\bf L}=\tilde{\bf {L}}^T$, where the tilde stands for the matrix when operating under the time-inverse protocol.  The breaking of the Onsager symmetry seems to have a number of puzzling consequences, in particular concerning the efficiency at maximum power and efficiency at minimum dissipation, which have been clarified recently \cite{benenti2011thermodynamic,shiraishi2016universal,proesmans2016power}. 

``Standard" thermodynamics deals with macroscopic systems. Since the very beginning, {questions have been} posed about its applicability to small-scale devices, with the Maxwell demon the most notable illustration. Over the past two decades, thermodynamics {has been extended to describe small systems, including their thermal fluctuations.}  The most striking result is that the second law---the positivity of entropy production---is replaced by a symmetry property for the probability distribution of this quantity, which reproduces the positivity for the average as a subsidiary consequence. The implications on the fluctuating efficiency of a small-scale engine were  elucidated more recently \cite{verley_unlikely_2014,verley_universal_2014,rana_single-particle_2014,rana2015anomalous,gingrich_efficiency_2014,esposito2015efficiency,polettini_efficiency_2015,proesmans2015efficiency,proesmans_stochastic_2015,pilgram2015quantum,Proesmansfcs,jiang2015efficiency,agarwalla2015full,1367-2630-17-9-095005,vroylandt2016efficiency,martinez2016brownian,dinis16,crepieux16,okada2016heat}.  One surprising conclusion is that the maximum efficiency, corresponding to a reversible operation, can occur by chance but is, for a time-symmetric operation, exponentially less likely than any other finite efficiency.  The specific form of the probability distribution for the efficiency (or, more precisely, its large deviation function) is, in the regime of linear response, fully specified in terms of the Onsager coefficients, providing a novel relation between fluctuation and dissipation.

The connection between Carnot efficiency and the second law puts a special emphasis on thermal machines. The second law, however, applies equally well to  isothermal energy transformations, in which one form of ``work"  is transformed into another form of ``work."  Such transformations are actually ubiquitous in biological systems, {for example} the notable role of ATP as energy converter in the cell \cite{alberts14,sugawa2016f1}.  In this case, efficiency is defined as output over input work.  Maximum efficiency is again reached for a reversible operation and is equal to $1$, expressing the thermodynamic possibility of a lossless work transformation. 


{The main purpose of this paper is to explicitly test and illustrate, both theoretically and experimentally, the aforementioned crucial features of stochastic thermodynamics on an exactly solvable isothermal Brownian engine, driven by a duo of time-periodic forces.  In particular, we  evaluate explicitly the Onsager matrix, describing the dissipation and interaction of these two forces, verify the Onsager-Casimir relation, and check the newly discovered connection between different operational regimes of the engine (maximum power, maximum efficiency, minimum dissipation).  Turning to the stochastic features, we verify the fluctuation theorem, evaluate the probability distribution of stochastic efficiency and the corresponding large deviation function, and check its signature features.  Last but not least, all these results are experimentally tested and reproduced without adjustable parameters on a colloidal system.}


\section{Linear irreversible thermodynamics}
\label{sec:linear-irreversible-thermodynamics}

%
We consider an open system in contact with one or several (ideal) reservoirs of heat, particles, or both, {which} can exchange work with one or several (ideal) work sources. The rate of change of its entropy has  two contributions, an {\textit{entropy flow}} term $\dot{S}_e$, representing the exchange of entropy with the environment, and an {\textit{entropy production}} contribution $\dot{S}_i$, describing the internal irreversible processes:
\begin{equation}\label{sl}
	\dot{S} = \dot{S}_e + \dot{S}_i \,.
\end{equation}
The second law requires that the  entropy production be non-negative:
\begin{equation}\label{ep}
	\dot{S}_i \geq 0 \,.
\end{equation}
Furthermore, the entropy production can, at a steady state, typically be written as a bilinear sum over the constitutive irreversible processes \cite{kondepudi2014modern,onsager1931reciprocal,onsager1931reciprocal2}:
\begin{align}
	\label{ep1}
	\dot{S}_i = \sum_n  J_n X_n \,.
\end{align}

\noindent {Here, the $X$'s represent, for each process, the thermodynamic forces quantifying the applied nonequilibrium constraint, for example a gradient in chemical potential.  The $J$'s represent  the corresponding fluxes, for example a work or particle flow.}  The second law stipulates that, {given} a single constraint $\dot{S}_i=J X$, flow and force have to be in the ``same direction," $J X\geq 0$ (for example, particles moving to lower chemical potential).  But with \textit{two flows}, one contribution to the entropy production can be negative, provided the overall entropy production is positive: 
\begin{align}
	\label{ep11}
	\dot{S}_i = \dot{S}_{i,1}+\dot{S}_{i,2} 
	\end{align}
has to be non-negative, but considering the constitutive processes
\begin{align}
	 \dot{S}_{i,1}=J_1 X_1 \;\;\;\;   \dot{S}_{i,2}=J_2 X_2\,
\end{align}
one can have a ``load" $\dot{S}_{i,1}=J_1 X_1 < 0$ if the ``drive" $\dot{S}_{i,2}=J_2 X_2 \geq -J_1 X_1$.
This observation corresponds, in fact, to the defining principle of an engine.  {In the best-known example, the heat engine, a ``downhill" flow of heat produces work via an ``uphill" flow against the other (e.g., mechanical, chemical or electrical) force.}  The principle, however, applies equally well to isothermal energy transformations, where one form of work is transformed into another form of work{---a ``work to work" engine.}

{From} the above discussion and {from Eq.}~\eqref{ep1} for the entropy production, {a natural} alternative definition of a thermodynamic efficiency $\bar{\eta}$ {is}   
\begin{equation}
	\bar{\eta}=-\frac{\dot{S}_{i,1}}{\dot{S}_{i,2}} = -\frac{J_1 X_1}{J_2 X_2} \leq 1 \,.
\label{Effdef}
\end{equation}
The universal upper bound  $1$ is reached for zero entropy production, i.e., for a reversible operation.  Note that this equilibrium point is also the point of flux reversal; i.e., an infinitesimal change of the forces at this point can switch the direction of both fluxes, {exchanging ``drive" and ``load."}  {The} definition of efficiency {is then} modified to ${\bar{\eta}} = {-J_2 X_2}/{J_1 X_1}$.

A further simplification arises by noting that in many  cases---chemical reactions being a notable exception---the forces are effectively weak (e.g., weak gradients). Since the fluxes vanish in the absence of the forces (which corresponds to an equilibrium state), {a Taylor expansion to lowest order in the forces gives}
\begin{equation}
\begin{aligned}
	J_1 &= L_{11} X_1+L_{1 2}X_2 \,, 	 \\[3pt]
	J_2 &= L_{21} X_1+L_{2 2}X_2 \,. 	
\label{Jdefs}
\end{aligned}
\end{equation}
These, together with Eqs.~\eqref{sl}--\eqref{ep1}, form the {core} equations of linear irreversible thermodynamics.  The second law now stipulates that the quadratic form 
\begin{equation}
	\dot{S}_i=L_{11} X_1^2+(L_{1 2}+L_{21}) X_1X_2+L_{2 2}X_2^2 \geq 0 \,,
\end{equation}
has to be non-negative, implying
\begin{equation}
	L_{11} \geq0 \,, \;\;\; L_{2 2}\geq 0 \,, \;\;\; (L_{12}+L_{21})^2 \leq 4 L_{11} L_{2 2} \,.
\end{equation}

Onsager discovered an {important additional} constraint on these coefficients \cite{onsager1931reciprocal,onsager1931reciprocal2}. He showed that, for properly defined fluxes and forces (and variables {that are} even under time reversal), the  matrix of the Onsager coefficients $\bold{L}$ {also} has to be symmetric, with $L_{12}=L_{21}$.  More generally, {microscopic} reversibility implies the  Onsager-Casimir symmetry $L_{12}=\tilde{L}_{21}$, where the tilde refers to inversion of the variables {that are} odd under time inversion \cite{casimir1945onsager} (for example, a velocity field \cite{perez1990motion} or a magnetic field \cite{degroot11,saito2008symmetry}). 
{Recently, it was realized that the Onsager-Casimir symmetry should also apply to variables that are \textit{even} under time reversal but are driven periodically by a \textit{time-asymmetric}  force.}
\cite{izumida2009onsager,izumida2010onsager,brandner2015thermodynamics,proesmans2015onsager,proesmans2015linear,brandner2016periodic,cerino2016linear}.  The latter situation {is} of great interest, as many engines operate in a time-asymmetric fashion.  For example, the time-reverse of a Carnot cycle no longer functions as an engine but rather as a refrigerator or heat pump!

Concerning the operational conditions of an engine, one may obviously be interested in maximizing the output, $P=-TF_1J_1$, {which is typically proportional to an output power and where $T$ is the temperature of the power-producing device.  But one may also be interested in maximizing the efficiency $\bar{\eta}$ or in minimizing the dissipation $\dot{S}_i$.}  The simplest way to perform such {optimizations} is to vary the load parameter $F_1$.  The optimization is, however, hampered by a trade-off between power and efficiency, which makes it impossible to optimize {both features} at the same time.  In fact, within the linear approximation of Eq.~\eqref{Jdefs}, straightforward algebra implies that power and efficiency in both regimes are linked \cite{proesmans2016power}:
\begin{equation}
	\bar{\eta}_{\rm MP} = \left( \frac{P_{\rm ME}}{2P_{\rm MP}-P_{\rm ME}} \right) \, 
		\bar{\eta}_{\rm ME} \,,
\end{equation}
where the subscripts MP and ME refer to the engine operating under maximum power and  maximum efficiency, respectively.  This relation further splits into two separate conditions when the Onsager symmetry is valid \cite{proesmans2016power,jiang2014thermodynamic,bauer2016optimal,ryabov2016maximum}:
\begin{equation}
	\frac{P_{\rm ME}}{P_{\rm MP}} = 1-\bar{\eta}_{\rm ME}^2 \,, \qquad
	\bar{\eta}_{\rm MP} = \frac{\bar{\eta}_{\rm ME}}{1+\bar{\eta}_{\rm ME}^2} \,.
\label{PED}
\end{equation}
Furthermore, these two regimes can also be linked to the limit of minimal dissipation (subscript mD):  
\begin{equation}
	P_{\rm mD} = 0 \,, \qquad 
	T \dot{S}_{\rm mD} = \left( \frac{1}{\bar{\eta}_{\rm MP}}-2 \right) P_{\rm MP} \,.
\label{PED2}
\end{equation}

\section{Gaussian stochastic irreversible thermodynamics}

The above discussion refers to macroscopic systems, with no reference whatsoever to fluctuations. {However, it turns out that a thermodynamic description that includes fluctuations away from a ``thermodynamic limit" leads to a much more profound and satisfying formulation of the second law.   This should not, in retrospect, be a surprise as, from the very beginning, Boltzmann himself stressed that the second law should be understood and interpreted in a statistical sense.}  We  briefly review the salient features of this stochastic thermodynamics \cite{seifert_stochastic_2012,van_den_broeck_ensemble_2014} in the context of a Gaussian approximation, which is relevant for the work to work engine introduced below.  The first observation to make is that the quantities observed in an experiment will fluctuate from one run to another.  We will denote such fluctuating quantities by the lower-case version of their macroscopic analogue; for example, the fluctuating work delivered by the engine is represented by $w$, the fluctuating flux by $j$, etc..
The stochastic entropy production for an engine thus reads (cf.~Eq.~\ref{ep1}):
\begin{equation}
\label{ep2}
	\dot{s}_i = \dot{s}_{i,1} + \dot{s}_{i,2} \,,
\end{equation}
with
\begin{equation}
	\dot{s}_{i,1} = j_1 X_1 \,, \quad  \dot{s}_{i,2} = j_1 X_2 \,.
\label{sep}
\end{equation}
Note that the thermodynamic forces are supposed here to be imposed by macroscopic constraints and hence are not fluctuating.  The second law is now replaced by a symmetry property, sometimes referred to as the fluctuation theorem. The quantities of interest  are the sample averages of the two contributions to the entropy production, measured over a  time $t$:
\begin{equation}\label{defsigma}
	\sigma_{1} = \frac{1}{t}\int^{t}_0dt\, \dot{s}_{i,1} \,, \quad
	\sigma_{2} = \frac{1}{t}\int^{t}_0dt\, \dot{s}_{i,2} \,.
\end{equation}

The fluctuation theorem (in its simple asymptotic form) imposes the following constraint:
\begin{equation}
\begin{aligned}
\label{sft}
	\frac{P(\sigma_1,\sigma_2)}{\tilde{P}(-\sigma_1,-\sigma_2)} &\sim 
		\exp{\left[\int^{t}_0dt\, \dot{s}_{i}/k_B  \right]} \,= 
		\exp{\left[ (\sigma_1+\sigma_2)t/k_B \right]} \,. 
\end{aligned}
\end{equation}
Here, ${\tilde{P}}$  denotes the probability distribution for the time-reversed experiment. The $\sim$ sign refers to the fact that the relation is valid for asymptotically large times.  (The result may be  valid for all times under supplementary conditions; see, e.g., \cite{becker_echo_2015}.)

To  study the implications of the fluctuation theorem, we consider the situation in which the fluctuations in the sample entropy fluxes $\sigma_1$ and $\sigma_2$ are described by a {bivariate} Gaussian distribution,  with averages $\left< \sigma_i \right>$ and covariance matrix $C_{ij} = \left< \sigma_i \,\sigma_j\right> - \left< \sigma_i\right> \left<\sigma_j \right>$:
\begin{equation}
	P(\sigma_1,\sigma_2)= \frac{1}{2\pi \sqrt{\det \bold{C}}} \, e^{-\frac{1}{2} 
	\sum_{i,j} \left( \sigma_i-\left< \sigma_j\right> \right) {C_{ij}^{-1}} \left( \sigma_j-\left<\sigma_j\right> \right) } \,.\label{pw1w2}
\end{equation}
This Gaussian ansatz  is typically valid for asymptotic long times and, even then, only as a first approximation of a large deviation function.  In the model that we discuss in the sequel, it is exact and valid for all times; hence, we have written an equality sign here.  
Combined with the fluctuation theorem, Eq.~\eqref{sft}, we derive fluctuation-dissipation relations between the Onsager  coefficients, which characterize the average response, and corresponding correlation functions (other conditions relate forward to time-reverse averages; see Appendix for more details):
\begin{equation}
	C_{ij} \sim k_B \,X_i  X_j \frac{\left(L_{ij}+L_{ji}\right)}{t} \,.
\label{CovOns}
\end{equation}
Within the Gaussian approximation Eq.~\eqref{pw1w2}, the stochastic thermodynamic properties of the engine are fully characterized in terms of the Onsager response coefficients.  In particular, one can discuss the novel issue of stochastic efficiency $\eta$:
\begin{equation}
\label{stocheta}
	\eta=-\frac{\sigma_1}{\sigma_2} \,.
\end{equation}
Its probability distribution, being the ratio of correlated Gaussian random variables, is given by \cite{polettini_efficiency_2015,Proesmansfcs,marsaglia2006ratios}:
\begin{widetext}
\begin{equation}
\begin{aligned}
	P_t(\eta)&=\int^{\infty}_{-\infty} d\sigma_1\, \int^{\infty}_{-\infty} d\sigma_2\, 
		\delta \left( \eta+\frac{\sigma_1}{\sigma_2}\right)p_t(\sigma_1,\sigma_2) \\[3pt]
	&= \frac{e^{c(\eta)}\sqrt{-[a(\eta)^2+2b(\eta)c(\eta)]}
		\left[ 2+\frac{\left| a(\eta) \right|}{\sqrt{b(\eta)}} e^{\frac{a(\eta)^2}{2b(\eta)}}\sqrt{2\pi} 
		\text{ erf} \left( \frac{\left| a(\eta) \right|}{\sqrt{2b(\eta)}} \right) \right]}
		{2b(\eta)\pi {\left| {\left<\sigma_1\right>}+\eta {\left<\sigma_2\right>} \right|}} \,,
\label{EfficiencyProb}
\end{aligned}
\end{equation}
\end{widetext}
with
\begin{equation}
\begin{aligned}
	a(\eta) &=\frac{C_{22} \eta \left<\sigma_1\right>-C_{11} \left<\sigma_2\right>+C_{12} 
		(\left<\sigma_1\right>-\eta \left<\sigma_2\right>)}{{\det \bold{C}}} \,, \\[3pt]
	b(\eta) &=\frac{C_{11}+2C_{12}\eta+C_{22}\eta^2}{{\det \bold{C}}} \,, \\[3pt]
	c(\eta) &=-\frac{C_{22}\left<\sigma_1\right>^2-2C_{12}\left<\sigma_1\right>\left<\sigma_2\right>+C_{11}
		\left<\sigma_2\right>^2}{2 \, {\det \bold{C}} } \,.
\label{EfficiencyProbCoefs}
\end{aligned}
\end{equation}
%
At this stage, the above expression for  $P_t(\eta)$ is a purely mathematical result, obtained from Eqs.~\eqref{pw1w2} and \eqref{stocheta}. 
To incorporate thermodynamic information, we consider the  large $t$ limit. $P_t(\eta)$ then assumes a large deviation form:
\begin{equation}
	{P_t}(\eta) \sim \exp{[-t\mathcal{J}(\eta)]} \,,
\end{equation}
characterized by the  \textit{large deviation function} $\mathcal{J}(\eta)$:
\begin{equation}
	\mathcal{J}(\eta) = -\lim_{t \rightarrow \infty} \frac{1}{t} \ln P_t(\eta) \,.
\label{eq:large-deviation-def}
\end{equation}
The latter can be calculated explicitly from Eq.~\eqref{EfficiencyProb} 
\cite{verley_unlikely_2014,polettini_efficiency_2015,jiang2015efficiency,Proesmansfcs}. 
One finds:

\begin{eqnarray}
\label{eq:large-deviation-explicit1}
	\mathcal{J}(\eta) 
	&=& -\lim_{{t \to \infty}} \frac{2b(\eta)c(\eta)+a(\eta)^2}{2b(\eta)t}  \nonumber \\
	&=&\lim_{t\rightarrow\infty} \frac{\left( \left< \sigma_1\right> 
		+ \eta \left< \sigma_2 \right>\right)^2}{2t\left(C_{11}+2\eta C_{12}+\eta^2C_{22}\right)} \,.
\end{eqnarray}

This expression can be further simplified by combining with the thermodynamic expressions for the averages  $\left( \sigma_i\right>=J_iX_i$, cf.~Eqs.~\eqref{Jdefs}, \eqref{sep} and \eqref{defsigma}, and the constraint imposed by the fluctuation theorem on the correlation function $C_{ij}$, cf. Eq.~\eqref{CovOns}. One thus obtains the following elegant expression in terms of response coefficients and thermodynamic forces: 
\begin{align}
	\mathcal{J}(\eta) =\frac{1}{4 k_B}
	\frac{\left( 
		\begin{bmatrix} X_1 & \eta X_2 \end{bmatrix} \boldsymbol{L}
		\begin{bmatrix} X_1 \\ X_2 \end{bmatrix} 
	\right)^2}
		{\begin{bmatrix} X_1 & \eta X_2 \end{bmatrix} \boldsymbol{L}
		\begin{bmatrix} X_1 \\ \eta X_2 \end{bmatrix}} \,.
\label{eq:large-deviation-explicit}
\end{align}
As can be verified by inspection (noting ${\bf J}=\boldsymbol{L}{\bf X}$), the large deviation function has a minimum equal to zero at the most probable value of the efficiency, $\mathcal{J}(\bar{\eta})=0$, with $\bar{\eta}$ given by Eq.~\eqref{Effdef}.  Any efficiency different from this  macroscopic efficiency becomes exponentially unlikely in the large time limit. A more surprising feature is that $\mathcal{J}(\eta)$ is a non-convex, non-monotonic function. It converges to  the single asymptotic value $J_2^2/(4k_B L_{22})$ for $\eta \rightarrow \pm \infty$ and displays  a single  other extremum, namely a maximum, at an intermediate value of $\eta$, which is larger than $\bar{\eta}$.  Furthermore, this value is located at reversible efficiency, $\eta_{\rm rev}=1$, if and only if the Onsager matrix satisfies $L_{12}=L_{21}$ \cite{jiang2015efficiency}.  

To complete the connection with cyclic thermodynamic small-scale machines, and in particular with the concrete model discussed in the sequel, we note that the above results remain valid for periodically driven systems, with all quantities replaced by their average over one period. Time $t$ is  replaced by the number of cycles $n$. Hence it will be sufficient to evaluate the corresponding Onsager coefficients to characterize the stochastic thermodynamic properties, at least within the Gaussian linear-response approximation expounded above.  In particular, we conclude that for a machine operating under a time-symmetric protocol, reversible efficiency becomes exponentially less likely than any other finite efficiency. We expect to see this signature appear as a pronounced minimum,
developing in the probability distribution ${P}_n(\eta)$ for the efficiency as the number of cycles $n$ is increased.

\section{Gaussian stochastic irreversible thermodynamics for a Brownian duet}

We consider a single overdamped Brownian particle moving in one dimension obeying {a Langevin equation:}
\begin{equation}
	\dot{z}(t) = -\gamma^{-1}\frac{\partial}{\partial z}U(z,t) + \sqrt{2D} \, \xi(t) \,.\label{Langevin}
\end{equation}
Here $\xi(t)$ is a Gaussian white noise:
\begin{equation}
	\left< \xi(t) \right> = 0, \qquad 
	\left< \xi(t) \, \xi(t')\right> = \delta(t-t') \,,
\end{equation}
where $\gamma$ is the friction coefficient, $D$ the diffusion constant, $T$ the temperature of the bath, and ${z}(t)$ the position of the particle.  The friction coefficient is linked to the diffusion coefficient by the Einstein relation $D~=~k_B T/\gamma$.

In order to confine the particle, one part of the potential $U(z)$ corresponds to  a static harmonic trapping potential $\kappa{ z}^2/2$.  In order to operate as an engine, we apply, in addition, a time-periodic forcing $F(t)$.  Before including the effect of the fluctuations, we {first describe}  the average behavior.

To do this, {consider a macroscopic} object with a single {(non-fluctuating)} degree of freedom $Z$ (e.g., {its position) in a potential} $U(Z) = \kappa Z^2/2$, with $\kappa$ the spring constant.  It undergoes overdamped motion in a heat bath at temperature $T$ subject to {an additional}  time-dependent external force $F(t)$.  From a thermodynamic point of view, the object with its spring energy is {an} open system.  For simplicity, we assume that it has no further ``internal structure" that is modified by the elongation of the {spring; hence,} the entropy of the system remains unchanged under elongation{---think of a particle in an externally applied parabolic potential rather than a particle attached to a rubber spring.}  The heat bath is supposed to be ideal; i.e., it exchanges heat in a reversible way and does not produce entropy on its own.  {If we neglect inertia, the equation of motion for the object is} 
\begin{equation}
\label{eqm}
	\dot{Z} = \gamma^{-1} \left[-\kappa{ Z} + F(t) \right] \,.
\end{equation}

Let us first consider  {``loading"} the spring.  This can be done reversibly, by applying a force infinitesimally larger than the {restoring} spring force.  {We set }$F(t) \approx \kappa Z(t)$.  The work done in bringing the object from an initial position $Z_i = 0$ {slowly} to a final position $Z_f$ is $W = \int_{0}^{Z_f} F(t) \, dZ(t) = \tfrac{1}{2} \kappa{ Z_f}^2$, which {equals exactly} the increase of internal energy of the spring.  {By the first law, no heat is exchanged with the bath, and $\Delta_eS=0$.  As the entropy of the system is unchanged under elongation, $\Delta S = \Delta_i S+\Delta_e S = 0$,} or $\Delta_i S = -\Delta_e S = 0$, as expected for a reversible process. 

{Alternatively, we can load the spring abruptly, by} switching on a constant force $F_0$ at time $t=0$. {We set} $F(t)=F_0 \, \theta(t)$, where $\theta$ is the Heaviside {step} function.  Under {the} influence of this force, the average particle position shifts from the initial position $Z_i=0$ to the final position $ Z_f=F_0/\kappa$. The work {is just force $\times$ displacement:}  $W = F_0 Z_f  = F_0^2/\kappa$. As before, the Brownian particle gains potential energy $\Delta E= \kappa{ Z_f}^2/2 = F_0^2 /(2\kappa) = W/2$, which is, however, less than the work that is performed.  {According} to the first law, $\Delta E=W+Q$, the heat towards the system $Q = \Delta E - W = - F_0^2/(2\kappa)$ is {then} negative.  An amount $-Q$ has thus been dissipated in the bath.  {Since} $\Delta S=\Delta_i S+\Delta_eS=0$, we conclude that $\Delta_i S=-\Delta_eS=Q/T = F_0^2/(2T\kappa)$ is positive, as it should be. 

For the above construction to operate as an engine, we need {to add} a resetting mechanism; i.e., the object needs to return to its original state. {We then need} to repeat the same operation over and over again, {which} can be achieved by considering a time-periodic driving $F(t)=F(t+\mathcal{T})$, with $\mathcal{T}$ the period of the driving. The equation of motion, Eq.~\eqref{eqm}, can still be solved exactly.  After an initial transient, the system will settle into a ``steady state" in the long-time limit $t\rightarrow \infty$:
\begin{equation}
\begin{aligned}
\label{zewq}
	Z(t) &= \frac{1}{\gamma} \int^{\infty}_0 d\tau \, e^{-\frac{k\tau}{\gamma}}F(t-\tau) \\
	\dot{E}(t) &= \kappa Z(t) \dot{Z}(t) \\
	\dot{W}(t) &= F(t) \dot{Z}(t) \\
	\dot{Q}(t) &=\dot{E}(t)-\dot{W}(t)=-\gamma \dot{Z}(t)^2 \,.
\end{aligned}
\end{equation}
A crucial observation to make is that this ``steady state" is time-periodic: $Z(t)$ inherits the periodicity of the driving  $Z(t)=Z(t+\mathcal{T})$, and so do the other variables. Hence it is quite natural to investigate the behavior of the system when {averaging} over one period $\mathcal{T}$.
It will be convenient to do so in terms of the  time-periodic driving written as
\begin{equation} 
	F(t) = F_0 \, g(t) \,,
\end{equation}
with $g(t)$ a time-periodic, dimensionless function $g(t)=g(t+\mathcal{T})$ {and $F_0$  a measure of the amplitude of the driving $F(t)$.}  

Returning to the thermodynamic picture, we first make the important observation that the  energy is the total derivative of the periodic function $\kappa Z^2(t)/2$; hence, its change over one period is identically zero.  Consequently, the system is merely dissipating, in the course of every period, the  input work into output heat.  The corresponding entropy production, averaged over one period (still denoting it, by a slight abuse of notation, as $\dot{S}_i$) is given by
\begin{eqnarray}
\label{Si0}
	\dot{S}_i &=&\; \frac{1}{\mathcal{T}} \int^{\mathcal{T}}_0 dt \, \frac{\dot{W}(t)}{{T}} 
		= \; -\frac{1}{\mathcal{T}} \int^{\mathcal{T}}_0 dt \, \frac{\dot{Q}(t)}{{T}} \nonumber\\
		&=& \frac{{\gamma}}{T\mathcal{T}} \int^{\mathcal{T}}_0 dt\,\dot{Z}(t)^2 \,.
\end{eqnarray}
One can now rewrite this result  as follows in the ``standard notation" of linear irreversible thermodynamics:
\begin{equation}
\label{epX}
	\dot{S}_i= XLX \,,
\end{equation}
with the (scalar) thermodynamic force defined as
\begin{equation}
	 X = \frac{F_0}{T} \,.
\end{equation}
The Onsager coefficient is most easily obtained by first replacing, in  Eq.~\eqref{Si0}, one factor $\dot{Z}$ by its expression from the equation of motion Eq~\eqref{eqm}.  Noting that the integral over one period of $Z \dot{Z}$ is zero and replacing the second factor $\dot{Z}$ by using  Eq.~\eqref{zewq}, one immediately  finds the following exact expression for the (scalar) Onsager coefficient $L$:
\begin{equation}
\begin{aligned}
\label{L1}
	L &= \frac{T}{\mathcal{T}\gamma}\int^{\mathcal{T}}_0dt\int^{\infty}_0d\tau\, 
		g(t) \, \dot{g}(t-\tau) \, e^{-\frac{k\tau}{\gamma}} \,.
\end{aligned}
\end{equation}

The second law implies that this coefficient should be non-negative.
This is most easily verified by using a Fourier series for the  function $g(t)$, periodic with frequency $\omega = 2\pi/{\mathcal{T}}$ :
\begin{equation}
	g(t) = \sum_{n=1}^{\infty} \left[ a_n \sin {\left(n\omega t\right)} + 
		b_n \cos{ \left( n\omega t\right) } \right] \,.
\end{equation}
 One finds (setting ${\omega'} = \omega t_r = \omega\gamma/\kappa$):

\begin{equation}
	{L = \frac{T}{2\gamma} \sum_{n=1}^{\infty} 
		\frac{{n^2 \omega'^2}}{{n^2 \omega'^2 +1}} \,
		\left( a_n^2+b_n^2 \right) {\ge 0} \,.}
\end{equation}
Note that the entropy production Eq.~\eqref{epX} can also be written as 
\begin{equation}
	\dot{S}_i = JX \qquad J = LX \,,
\end{equation}
with $J$ the (scalar) flux.
 
The above dissipative machine appears  to be utterly uninteresting.  It is, however, clear from Sec.~\ref{sec:linear-irreversible-thermodynamics} how to transform  it into a genuine engine by applying a periodic modulation that is the {\it sum} of two separate contributions, $F(t)=F_1(t)+F_2(t)$.  {Here, $F_1(t)=F_1(t+\mathcal{T})=F_{1,0}g_1(t)$ plays} the role of the load, and  $F_2(t)=F_2(t+\mathcal{T})=F_{2,0}g_2(t)$ the role of the {drive}.  {The two} forces could be mechanical, electrical, chemical, or {some mixture} but need not be further specified.  Eqs.~\eqref{zewq} remain valid with the understanding that the work rate can be split into two separate contributions, coming from $F_1$ and $F_2$, respectively:

\begin{equation}
\begin{aligned}
	\dot{W}(t) &= \dot{W}_1(t) +\dot{W}_2(t)  \\
	\dot{W}_1(t) &= F_1(t)\dot{Z}(t) \,,  \qquad \dot{W}_2(t) = F_2(t)\dot{Z}(t)  \,.
\end{aligned}
\end{equation}
Similarly the entropy production, averaged over one period, (cf. Eq~\ref{Si0}), can be written as follows (cf. Eq.~\ref{ep11}):
\begin{equation}
\begin{aligned}
\label{Si1}
	\dot{S}_i &= \dot{S}_{i,1}+\dot{S}_{i,2} \\[3pt]
	\dot{S}_{i,1}&=\; \frac{1}{\mathcal{T}} \int^{\mathcal{T}}_0 dt \, 
	\frac{\dot{W}_1(t)}{{T}},\;\;\;\dot{S}_{i,2}=\; \frac{1}{\mathcal{T}} 
	\int^{\mathcal{T}}_0 dt \, \frac{\dot{W}_2(t)}{{T}}   \,.
\end{aligned}
\end{equation}

Alternatively, the entropy production can be expressed in terms of vectorial forces $ {\bf X}$, fluxes  $ {\bf J}$, and the Onsager  matrix, $\mathbf{L}$:
\begin{equation}
	\dot{S}_i  =  {\bf X J}={\bf XLX}\,, \qquad {\bf X}=\frac{{\bf F}_0}{T}\,, \qquad {\bf J} =\mathbf{L} {\bf X}\,.
\end{equation}
${\bf F}_0=(F_{1,0},F_{2,0})$ denotes the amplitude of the perturbations.  The components of the $2 \times 2$ Onsager matrix $\mathbf{L}$ can be read off from Eq.~\eqref{L1}:
\begin{align}
	L_{ij} = \frac{T}{\mathcal{T}\gamma} \int^{\mathcal{T}}_0 dt \int^{\infty}_0 d\tau \,
		g_i(t) \, \dot{g}_j(t-\tau) \, e^{-\frac{k\tau}{\gamma}} \,.
\label{OnsDef}
\end{align}
The macroscopic efficiency of the work to work conversion is given by
\begin{equation}
	\bar{\eta}= -\frac{\dot{S}_{i,1}}{\dot{S}_{i,2}} = -\frac{\frac{1}{\mathcal{T}}
	\int^{\mathcal{T}}_0dt\, \dot{W}_1(t)}{ \frac{1}{\mathcal{T}}
	\int^{\mathcal{T}}_0dt\,\dot{W}_2(t)} = 
	-\frac{X_1(L_{11}X_1+L_{12}X_2)}{X_2(L_{21}X_1+L_{22}X_2)} \,.
\end{equation}

We finally return to the Langevin description, Eq.~\eqref{Langevin}, and consider its stochastic thermodynamics operating as a work to work transforming engine subject to the time-periodic forces $F(t)=F_1(t)+F_2(t)$, operating in a confining harmonic potential.  We can repeat the derivations given above in terms of the stochastic thermodynamic quantities, which we identified by a lower-case notation. Focusing on the issue of stochastic efficiency, we make the crucial observation that the stochastic work rates $\dot{w}_i(t) = F_i(t) \dot{z}(t)$ are correlated Gaussian random variables, and hence so are the sample averages of the stochastic  entropy productions:
\begin{equation}
\begin{aligned}\label{sepm}
	\sigma_i = \frac{1}{n\mathcal{T}} \int^{n\mathcal{T}}_0dt\, \frac{F_i(t)\dot{z}(t)}{T}.\\
\end{aligned}
\end{equation}
We can thus copy the conclusions from the previous section concerning the stochastic efficiency:
\begin{equation}
	\eta = -\frac{\sigma_1}{\sigma_2} = -\frac{\frac{1}{n\mathcal{T}}
	\int^{n\mathcal{T}}_0dt\,\dot{w}_1(t)}{ \frac{1}{n\mathcal{T}}
	\int^{n\mathcal{T}}_0dt\,\dot{w}_2(t)} \,.
\label{eq:efficiency}
\end{equation}
In view of the exact Gaussian nature of the sample entropy productions, cf. Eq.~\eqref{sepm}, the results Eqs.~\eqref{EfficiencyProb}--\eqref{eq:large-deviation-explicit1} apply for any number of cycles $n$, while  Eq.~\eqref{eq:large-deviation-explicit} only applies  in the large-$n$ limit.  The Onsager coefficients are given in Eq.~\eqref{OnsDef}. As one can verify explicitly, they are related to the correlation functions by Eq.~\eqref{CovOns}, as should be the case for a bona fide physical model that obeys the fluctuation theorem.

%
%
The above, a ``spring duet," corresponds arguably to the simplest possible thermodynamic engine, in which a particle in a quadratic potential functions as the engine transforming a Gaussian stochastic input work $w_1$ into a Gaussian stochastic output work $w_2$.  Both standard and stochastic thermodynamics are fully described in terms of Onsager coefficients that characterise the average response properties per cycle.  

An interesting novel feature is that, by virtue of the time-periodic nature of the perturbation, the Onsager matrix need not be symmetric. In fact, the matrix displays the aforementioned Onsager-Casimir symmetry $L_{12}=\tilde{L}_{21}$, where the tilde refers to the same set-up but under time reversed drivings $\tilde{F}(t)=F({\mathcal{T}} - t)$.  This Onsager-Casimir symmetry can be easily seen if the two driving forces, $F_1(t)$ and $F_2(t)$, only differ by a phase. As a general shift in time of the driving forces does not alter the average entropy production, it is clear that the amount of dissipation in the absence of $F_1(t)$ would be the same as the amount of dissipation in the absence of $F_2(t)$, i.e.~$L_{11}=L_{22}$.  On the other hand, the system under study would be exactly the same (up to a shift in time), if the driving were time inverted and $F_1(t)$ and $F_2(t)$ interchanged, which means that $X_1 J_1=X_2\tilde{J}_2$.  Combining these two results immediately gives $L_{12}=\tilde{L}_{21}$.

\section{Experimental tests}
In order to test these ideas experimentally, we use a micron-scale colloidal particle in a feedback trap \cite{cohen05b,cohen05d,Jun14}.  The feedback trap is designed around a microscope that includes a camera read out and a sample cell with attached electrodes.  The sample cell is filled with deionized water and a solution of silica beads.  In such traps, the particle is freely diffusing but subject to controllable electric forces.  There is no true potential but rather a \textit{virtual potential} that is imposed by the feedback loop.  In brief, the experiment rapidly and repeatedly executes the following sequence:
\begin{itemize}
\item A camera images the particle. 
\item A computer algorithm identifies the particle and determines its position.
\item The force corresponding to the real potential is calculated.
\item The force is output by setting the appropriate electrode voltage.
\end{itemize}
The last step is the most delicate, requiring essentially continuous calibration in order to combat drifts over the hours-long duration of the experiments \cite{gavrilov14}.  Another requirement is that the feedback loop cycle time $t_s$ be short compared to the relaxation time $t_r = \gamma / \kappa$ \cite{jun12} In these experiments, $t_s = 0.005$ sec, while the relaxation time $t_r$ is of the order of a few $100$ milliseconds.

Experiments are done on a set of nominally identical but, in reality, slightly different colloidal particles.  To allow combining data sets from different particles, we will present the experimental results in terms of dimensionless parameters.  We thus define ``natural" scales for time $t_r$, length $x_r$, and force $F_r$, as a function of the diffusion coefficient $D$, the friction coefficient $\gamma$, and the stiffness of the potential $\kappa$.  These are given by
\begin{equation}
	t_r=\frac{\gamma}{\kappa} \,, \quad z_r=\sqrt{Dt_r} \,, \quad F_r=\sqrt{Dt_r} \, \kappa \,.
\label{eq:scales}
\end{equation}
In terms of these quantities, we also define a natural energy scale $E_r = z_r \, F_r = k_B T$, using the Einstein relation.  {Then, using primes} for scaled, dimensionless quantities, we have
\begin{align}
	t' = \frac{t}{t_r} \,, \quad \mathcal{T}' = \frac{\mathcal{T}}{t_r} \,, 
		\quad z' =  \frac{z}{z_r} \,, \quad F_{i,0}' = \frac{F_{i,0}}{F_r} \,.
\end{align}
The {scaled} equation of motion no longer depends explicitly on the experimental parameters.  
{
\begin{equation}
	\dot{z}'(t') = -z'(t')-F_{1,0}' \, g_1(t') - F_{2,0}' \, g_2(t') + \sqrt{2} R(t') \,.
\end{equation}
}

\subsection{Onsager coefficients:  Experimental analysis}

\begin{figure}[h!]
  \centering
 \includegraphics[width=200pt]{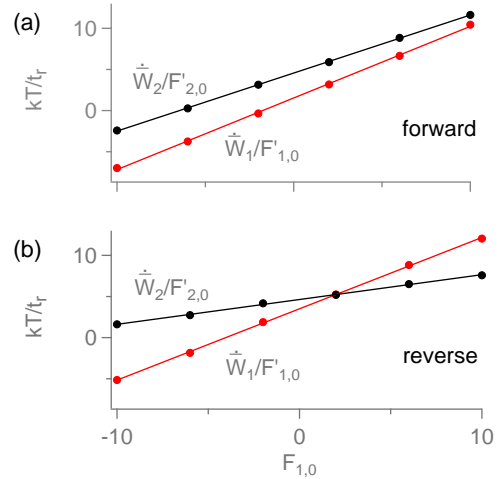}
 \caption{Experimentally determined $\dot{\bar{W_i}}/F_{i,0}'$, {$i=1,2$}, for the forward (a) and time-reversed (b) processes. Typical errors are around $\pm0.1~k_BT{/t_r}$.  Error bars are not shown,  as they are smaller than the markers indicating the mean values of the data.  Solid lines are least-squares linear fits.}
\label{OnsCoef}
\end{figure}

The first set of experiments tested Onsager-Casimir symmetry.  For the time-forward driving, we set 
\begin{equation}
\begin{aligned}
	g_1(t') &= \sqrt{2}\sin\left({\omega' t'}\right) \,, \\
	g_2(t') &= \sqrt{\frac{2}{3}} \left[ \sin \left( {\omega' t'} \right)
		+ \cos \left( {\omega' t'}\right)
		+ \cos \left( {2\omega' t'}\right) \right] \,.
\end{aligned}
\end{equation}
The time-inverted process is obtained by reversing the time-dependency of the driving.  Let us now fix some parameters:
\begin{equation}
\begin{aligned}
	\kappa =&~\frac{k_BT}{0.4D\cdot 1 {\rm ~sec}},\quad \mathcal{T}' =2.5  \\[3pt]
	 F_{i,0} = &~F_{i,0}'\cdot k_BT /  
		\sqrt{0.4D\cdot 1 {\rm ~sec}} \,.
\end{aligned}
\end{equation}
The diffusion coefficient was not known a priori and was measured during the experiment.  In dimensional units, one cycle {takes} $1$ second.  We set $F_{2,0}'=5$ and varied $F_{1,0}'$ between $-10$ and $10$.  We measured $100$ cycles for each value of $F_{1,0}'$.  The experimentally determined positions $z'(t')$ can now be used to calculate the work rates $\dot{\bar{W_i}} = \frac{1}{\mathcal{T}'}\int_0^\mathcal{T'} dt' \,  \dot{F}_i'(t') z'(t')$ for each cycle.  As the feedback loop operates in discrete time steps, the integral becomes the sum
\begin{align}
	\dot{\bar{W_i}} = \frac{1}{\mathcal{T}'} \sum_{k=1}^N \left( F_{i,k+1}'- F_{i,k}'\right) 		\frac{z_{k+1}' + z_k'}{2} \,,
 \end{align}
where $N$ is the number of time steps per cycle.

Figure~\ref{OnsCoef} plots $\dot{\bar{W}}_1/(F_{1,0}') = L_{11} F_{1,0}'+L_{12}F_{2,0}'$ and $\dot{\bar{W}}_2/F_{2,0}' = L_{21}F_{1,0}' + L_{22}F_{2,0}'$ as a function of $F_{1,0}'$ and shows linear least-squares fits to the result, whose slope and intercept determine the dimensionless Onsager coefficients.  We can do the same for the time-reversed process.  

\begin{table}[h!]
	\centering
	\caption{Experimentally determined Onsager coefficients for forward and reversed protocols, with theoretical predictions.}
	\label{table:onsagerCoefs}
	\begin{tabular}{|c|c|c|c|}
	\hline
	{index} & $\tilde{L}^T$& $L$ & $L_{\rm analytical}$ \\ \hline 
	11	& $~0.864\pm 0.006~$ 	& $~0.87\pm 0.01~$ 			& 0.863  \\ \hline
	12	& $~0.302 \pm 0.006~$ 	& $~0.30 \pm 0.01~$ 		& 0.300  \\ \hline
	21	& $~0.701 \pm 0.009~$  	& $~0.705\pm 0.004~$ 		& 0.697  \\ \hline
	22	& $~0.931\pm 0.009~$ 	& $~0.913\pm 0.006~$		& 0.896  \\ \hline	
	\end{tabular}
\end{table}

Table~\ref{table:onsagerCoefs} compares the experimental results with the analytical predictions from Eq.~\eqref{OnsDef}.  We see that the Onsager-Casimir relations, $\mathbf{L}~=~\tilde{\mathbf{L}}^T$, are indeed verified  to high precision, and so are the second-law constraints:  $L_{11} > 0$, $L_{22} > 0$, and $4L_{11}L_{22} \, \approx 3.18 \geq \left(L_{12}+L_{21}\right)^2 \approx 1.01$.  The small ($\lesssim 5$\%) but statistically significant differences between experimental results and analytical theory may reflect the finiteness of the time steps used in the feedback loop or other minor systematic experimental errors.

\begin{figure}[h!]
  \centering
 \includegraphics[width=200pt]{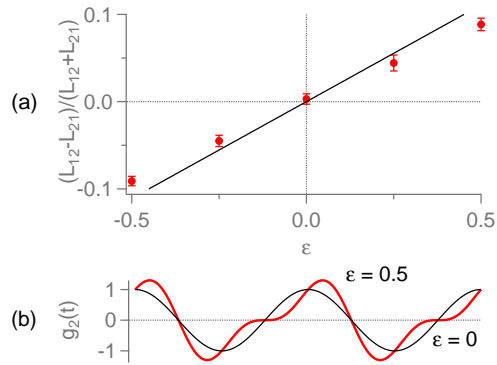}
 \caption{Asymmetry of Onsager coefficients.  (a)  $(L_{12}-L_{21}) / (L_{12}+L_{21})$, vs. temporal asymmetry.  {The red markers denote the experimental results, the solid black line  the theoretical curve.}  $\mathcal{T}'=2.5$ and $F_{2,0}'=5$.  $F_{1,0}'$ varies from $-10$ to $+10$ for each $\varepsilon$.  Work rates $\dot{\overline{W}}_i$ are calculated from 150 cycles.  (b) Illustration of the driving function $g_2(t')$, for $\varepsilon=0.5$ (red, asymmetric), compared to the symmetric case, $\varepsilon =0$ (black, symmetric).} 
\label{fig:SymmetryBreaking}
\end{figure}

Next, we show explicitly how breaking time-reversal symmetry of the driving function leads to an asymmetry of the Onsager response matrix $\mathbf{L}$.   The driving functions are 
\begin{equation}
\begin{aligned}
	g_1(t') &= \cos \left( \omega' t' \right) + \cos \left( 2\omega' t' \right) \,, \\
	g_2(t') &= \cos \left( \omega' t' \right) + \varepsilon \sin \left( 2\omega' t' \right) \,.
\label{eq:driving-funcs-sym-break}
\end{aligned}
\end{equation}
In Fig.~\ref{fig:SymmetryBreaking}(a), we report the asymmetry, $(L_{12}-L_{21}) / (L_{12}+L_{21})$, as a function of $\varepsilon$, which characterizes the amount of time-reversal symmetry breaking in $g_2(t')$, illustrated in Fig.~\ref{fig:SymmetryBreaking}(b). Classical Onsager symmetry, $L_{12}=L_{21}$, is restored if $\epsilon=0$.  As in Table~\ref{table:onsagerCoefs}, the difference between the slope of the experimental results and that of the theoretical prediction results from the finite update time $t_s$ of the feedback loop.

\subsection{Power-Efficiency-Dissipation relations}
In a second set of experiments, we tested the power-efficiency-dissipation relations, Eqs.~\eqref{PED}--\eqref{PED2} {using the time-symmetric driving functions}
\begin{equation}
	\begin{aligned}
	g_1(t') &= \cos \left( {\omega' t'} \right) \,, \\ 
	g_2(t') &= \cos \left( {\omega' t'} \right) 
		+ \frac{1}{8} \cos \left({4\omega't'} \right) \,.
	\end{aligned}
\end{equation}
We set $\mathcal{T}'=2.5$, $F_{2,0}'=5$ and  varied $F_{1,0}'$ between $-5.5$ and $0$, {taking $700$ cycles of data per value of $F_{1,0}'$}.  

\begin{figure}[h!]
	\centering
	\includegraphics[width=200pt]{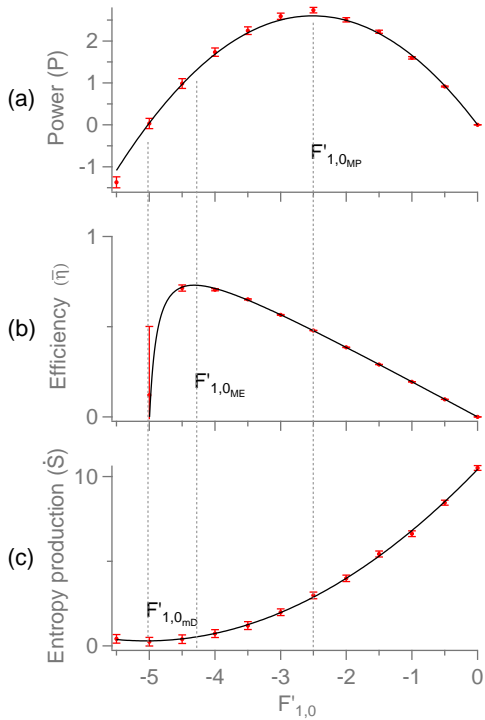}
	\caption{Power (a), efficiency (b) and entropy production (c) as a function of $F'_1$. The solid lines are least-squares fits. 
\label{PEDR}}
\end{figure}

{Figure~\ref{PEDR} shows fits to the power, entropy production, and efficiency.  From the extrema, we find}
\begin{equation}
	\begin{aligned}
	P_{\rm MP} &= 2.68\pm 0.03 &\bar{\eta}_{\rm MP} &= 0.477\pm 0.02  \\
	P_{\rm ME} &= 1.3\pm 0.2  &\bar{\eta}_{\rm ME} &= 0.73\pm 0.03   \\P_{\rm mD} &= 0.2\pm 0.2
	&S_{\rm mD} &= 0.29\pm 0.05 \,.
	\end{aligned}
\end{equation}
With these results, we can verify the power-efficiency-dissipation relations, Eqs.~\eqref{PED}--\eqref{PED2}:
\begin{equation}
	\begin{aligned}
	\frac{P_{\rm ME}}{P_{\rm MP}} &= 0.48\pm 0.08 \\
		& \approx 1-\bar{\eta}_{\rm ME}^2=0.47\pm 0.05 \,, 
		 \\[3pt]
	\frac{\bar{\eta}_{\rm ME}}{1+\bar{\eta}^2_{\rm ME}} &= 0.476\pm 0.006\approx \bar{\eta}_{\rm MP} \,, 
		 \\[3pt]
	\left( \frac{1}{\bar{\eta}_{\rm MP}}-2\right) P_{\rm MP} &= 0.3\pm 0.1\approx \dot{S}_{\rm mD} \,, 
		 \\[3pt]
	P_{\rm mD} &\approx 0 \,.
	\end{aligned}
\end{equation}

\subsection{Efficiency fluctuations}

We shall now study the efficiency fluctuations under time-symmetric driving.  In this case, the large deviation function should have a maximum at the reversible efficiency $\eta_{\rm rev}=1$, {implying a corresponding local minimum (dip) in the probability distribution at $\eta_{\rm rev}=1$ that emerges} for larger times.

For the driving, we set
\begin{equation}
\begin{aligned}
	g_1(t') &= \sqrt{2}\cos\left({\omega' t'}\right) \,, \\
	g_2(t') &= \frac{1}{\sqrt{13}} \left[ -5\cos \left( {\omega t'} \right)
		+ \cos \left( {4\omega' t'}\right) \right] \,.
\end{aligned}
\end{equation}
Furthermore, we set $\mathcal{T}' = 1.25$, $F_{1,0}' = F_{2,0}'= 10$ and $\kappa = k_BT/(0.8D \cdot 1$ sec.).  In dimensional units, $D = 0.25~\mu$m$^2/$sec., $\gamma=k_BT/D$, $\tilde{\mathcal{T}} = 1$ sec., and $F_{1,0} = F_{2,0} = 10 \sqrt{5} \, k_BT/\mu$m.  Following Eq.~\eqref{eq:efficiency}, the efficiency is defined as $\eta = -\left( 
	\int^{n\mathcal{T}'}_0 dt' \, \dot{w}_1(t') \right) / \left( 
	\int^{n\mathcal{T}'}_0 dt' \, \dot{w}_2(t') \right)$.  We can look at the probability distributions for {$n$= 1, 2, and 4 cycles.  From these three distributions, we can calculate the large deviation function} via the extrapolation procedure described in Appendix A of Ref.~\cite{Proesmansfcs}.  For the experimental analysis, we use the same data, which includes $78 \, 000$ cycles,  for the different distributions, by just looking at the different cycles separately or by measuring the efficiency over multiple cycles.

\begin{figure}[h!]
	\centering
	\includegraphics[width=200pt]{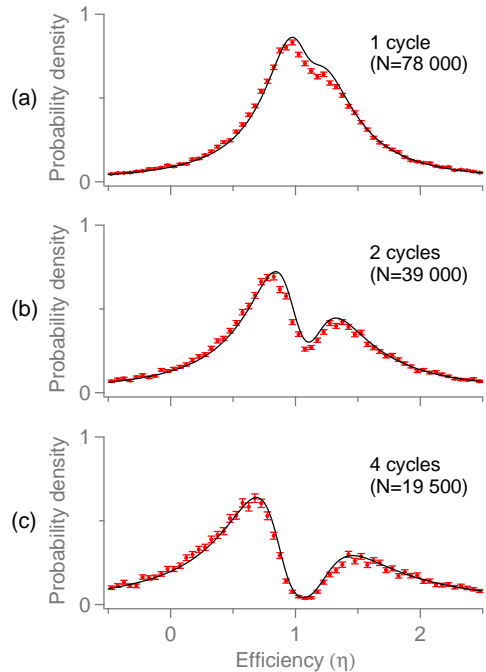}
	\caption{{Efficiency distributions for symmetric driving.  The solid lines show the  theoretical probability densities, calculated---not fit---from Eqs.~\eqref{EfficiencyProb} and \eqref{EfficiencyProbCoefs}.}
\label{EffDis}}
\end{figure}

\begin{figure}[h!]
	\centering
	\includegraphics[width=200pt]{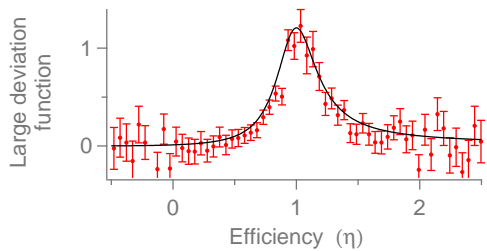}
	\caption{Large deviation function for the efficiency $\mathcal{J}(\eta)$, defined in Eq.~\eqref{eq:large-deviation-def}, for symmetric driving.  Solid line is calculated using Eq.~\eqref{eq:large-deviation-explicit}.}
\label{EffLDFS}
\end{figure}
Figure \ref{EffDis} shows the efficiency probability distributions when calculated by averaging the work values over $n =$ 1, 2, and 4 cycles.  The probability distributions clearly agree with the Gaussian  predictions, showing a local minimum at $\eta \approx 1$.  {Figure~\ref{EffLDFS} shows the measured large deviation function for efficiency, along with the curve calculated, with no adjustable parameters, from Eq.~\eqref{eq:large-deviation-explicit}.  The maximum at reversible efficiency, $\eta=1$, and the agreement with Gaussian predictions are clearly visible.}

\begin{figure}[h!]
	\centering
	\includegraphics[width=200pt]{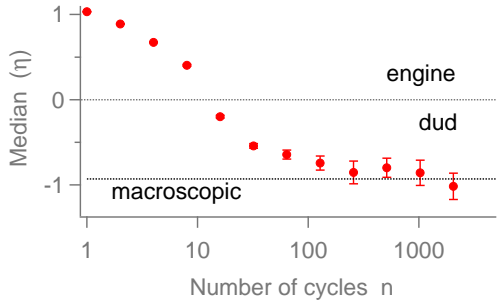}
	\caption{Median of the efficiency distribution as a function of the number of cycles $n$.  Dotted line indicates macroscopic efficiency, $\bar{\eta} \approx -0.93$.  Error bars estimated via the bootstrap method \cite{press07}.}
\label{fig:median}
\end{figure}

Although investigating directly the efficiency distribution $P_n(\eta)$ for a large number of cycles $n$ would require more data than we can collect from a single particle, we can use summary statistics instead.  Since $P_n(\eta) \to \eta^{-2}$ for large $|\eta|$, the average is not defined; however, the median is an appropriate, robust statistic to use for such fat-tailed distributions \cite{hanson96}.  Figure~\ref{fig:median} shows the median as a function of the number of cycles $n$.  Surprisingly, its value, while positive for small $n$ (as obvious from Fig.~\ref{EffDis}), becomes negative for $n \gtrsim 10$ cycles.  In the $n \to \infty$ limit, it converges to a macroscopic value that, from simulations, is estimated to be $\bar{\eta} \approx -0.93$.  One could classify the performance regime as a \textit{tease}:  it acts as an \textit{engine} (typical $\eta > 0$) when its performance is evaluated over short time intervals but as a \textit{dud} (typical $\eta < 0$) over longer time intervals.

Why is the behavior evaluated at small cycle number so different from the long-time, macroscopic limit?   In the small-$n$ limit, there are
large fluctuations that guarantee that the average work production is always negative.  As we use longer time intervals to calculate $\eta$, more and more negative fluctuations are included in the ``typical" value, which pulls its location steadily towards the macroscopic value.  The convergence to this macroscopic value is very slow, as the large deviation function has very small asymptotes: \begin{equation}
	\lim_{\eta\rightarrow\pm\infty}J(\eta)=0.013 \,,
\end{equation}
which in turn is a consequence of the fact that the engine operates near the so-called ``singular coupling" limit \cite{polettini_efficiency_2015}.

\section{Discussion and Conclusion}

We have seen that Brownian duets are a simple setting that is easy to analyze both theoretically and experimentally.  Their importance stems from their role in raising and illustrating important, fundamental issues in thermodynamics.  The second law of thermodynamics is typically derived via the discussion of the Carnot cycle.  An explicit exact calculation of this set-up is, however, possible only for a macroscopic system operated with an ideal gas in the reversible regime. Reversibility conveys to the construction the crucial property that it can function both as a thermal engine and as a refrigerator.  Combined with the assumed impossibility of a perpetuum mobile of the second kind, one can hence derive the famous inequalities, i.e., that efficiencies must be smaller than the Carnot efficiency and the ``famous" positivity of entropy production.  

While this derivation is one of the most beautiful in science, its drawbacks and shortcomings raise several fundamental questions.  As an example of a technical issue, how can an ideal gas stay at equilibrium while expanding adiabatically in the absence of a heat bath?  More fundamentally, the macroscopic limit, in which fluctuations are  ignored, obscures  the statistical nature of the second law and even conveys a belief---still broadly held---that the second law is a property of macroscopic systems alone.  

Following the footsteps of Boltzmann, nonequilibrium statistical mechanics has partially bridged some of the gaps in the derivation and understanding of the second law.  The basic connection between fluctuation and dissipation is core to the famous Onsager symmetry, fluctuation-dissipation theorem, and Green-Kubo relations.  These derivations, however, make no direct connection to thermodynamic machines.  Furthermore, they often require a somewhat disturbing mix-up between the micro and macro worlds.  For example, to connect with the world of equilibrium fluctuations, Onsager assumes the regression hypothesis, stating that macroscopic disturbances regress on average in the same way as microscopic deviations caused by spontaneous fluctuations.  Conversely, the derivation of microscopic linear response  \`a la Green Kubo has been justly criticized in its application to linear response for macroscopic laws \cite{vankampen71,vanvliet88}.
   
The present work proceeds from and partially completes the work of Einstein, Langevin, and Smoluchowski, by focusing on the statistical physics at the mesoscopic level of Brownian particles.  Both underdamped and the overdamped description of Brownian motion considered here have served well as an exemplar for nonequilibrium statistical physics.  By introducing the duet of two periodic forces acting on a Brownian particle, we connect with classic discussions of thermodynamic machines~\footnote{
Engine aspects of Brownian motion have been discussed in the context of Brownian motors, as, for example, in the review \cite{reimann02}.  The rectification observed here is, however, a purely nonlinear effect; hence, there is no simple connection with Onsager coefficients.  For a review in this context focusing more on thermodynamic considerations, see \cite{parrondo02}
}.  
Stochastic thermodynamics, with the second law replaced by a symmetry property for the probability distribution of the entropy production, provides the key ingredients to analyze this novel tale of efficiency. The intimate relation between fluctuation and dissipation  is brought into full light, as all aspects of this engine, including the telltale properties of its fluctuating efficiency, can be expressed in terms of the (linear response) Onsager coefficients.

The Brownian duet model has further advantages:  It can be solved in full analytic detail. Its experimental  implementation, including the gathering of sufficient statistics and its possible  technological applications, pose no special challenges.  Compared to the Carnot-engine construction, it does miss the heat-to-work transformation, as well as some aspects of stochastic thermodynamics that have already been illustrated by a mesoscopic Stirling engine involving a single Brownian particle in a breathing harmonic potential \cite{blickle12}.  However, non-isothermal conditions are not natural in the Brownian world---in the Stirling-engine experiment,  heating requires a strong laser pulse---nor, in fact, in most of the biological world.  Furthermore, applying stochastic thermodynamics to a heat-to-work transformation is more involved because of the non-Gaussian distribution of heat fluctuations \cite{farrago02,vanzon03, visco06,fogedby11}, rendering the theoretical analysis and interpretation of the engine and its efficiency much more complicated \cite{park16}.  

In any case, in this article, we have presented the first experimental observation of the dependence of the symmetry of Onsager response coefficients on that of the driving force; of the remarkable relations between power, efficiency, and dissipation that hold in the linear-response regime; and of the maximum of the large deviation function at reversible efficiency, a prediction that is ultimately a consequence of microscopic reversibility.  We also observe that the Brownian duet can function in a ``tease" mode:  when examined at short time intervals, it acts as a motor in the sense that a typical efficiency is positive, but when examined at longer time intervals, it acts as a dud.  These tests illustrate the power of stochastic thermodynamics in extending a subject, developed in the 19th century to describe macroscopic heat engines, to the mesoscopic, fluctuating world that is so much the focus of science in the 21st century.

\acknowledgments  We thank Laila Singh for suggesting the term ``tease."  This work was supported by the Flemish Science Foundation (FWO-Vlaanderen) travel grant K201516N and by a Discovery Grant to JB from NSERC (Canada).

\appendix*

\section{Covariance and Onsager matrix}
Inserting the expressions for the probability distribution for the entropy production in the forward process,  Eq.~\eqref{pw1w2}, and the one for the time-reversed process (the superscript tilde $\sim$  referring to the  time-reversed quantities),
\begin{multline}
	\tilde{P}(\sigma_1,\sigma_2) \\= \frac{1}{2\pi\sqrt{\det{\tilde{C}}}}
	\exp \left( -\frac{1}{2}\sum_{i,j} \left(\sigma_i - \left<\tilde{\sigma}_i\right>\right) 
	\tilde{C}^{-1}_{ij} \left( \sigma_j - \left<\tilde{\sigma}_j \right> \right)\right) \,,
\end{multline}
into the fluctuation theorem, Eq.~\eqref{sft} implies that the following equality should hold:
\begin{align}
	\left(\boldsymbol{\sigma}-\left<\boldsymbol{\sigma}\right>\right)C^{-1} 
		&\left(\boldsymbol{\sigma}
	-\left< \boldsymbol{\sigma} \right> \right) - \left( -\boldsymbol{\sigma} 
	- \left< \tilde{\boldsymbol{\sigma}} \right> \right) \tilde{C}^{-1}
	\left( -\sigma-\left< \tilde{\boldsymbol{\sigma}} \right> \right) \nonumber \\ 
		&+ \log\frac{\det C}{\det \tilde{C}} = -\frac{2t}{k_B} \bf{1}\boldsymbol{\sigma} \,.
\end{align}
Here, $\boldsymbol{\sigma}$ is the vector with components $\sigma_1,\sigma_2$ and $\bf{1}$ the vector with components $1,1$.   Note that we are considering the sample average entropy productions over a time $t$.  Their average and correlation function are thus a function of this time.  Since we are using the time-asymptotic form of the fluctuation theorem, the above equality, as the ones that follow below, should always be understood in the sense of a long-time limit.  The former equation should be valid for any $\boldsymbol{\sigma}$; hence, we can equate separately the zeroth-, first-,  and second-order terms of this quantity.  For the second order, we find 
\begin{equation}
	\boldsymbol{\sigma}C^{-1}\boldsymbol{\sigma} = \boldsymbol{\sigma}\tilde{C}^{-1} \boldsymbol{\sigma} \,,
\end{equation}
which has to be valid for all values of $ \boldsymbol{\sigma}$. Since $C$ and $\tilde{C}$ cannot differ by an anti-symmetric part as both matrices are fully symmetric, we conclude that
\begin{equation}
	C=\tilde{C} \,.
\end{equation}
As a result, $\log \bigl( \det C / \det \tilde{C} \bigr)$ is identically zero.
The first-order term in $\boldsymbol{\sigma}$ gives
\begin{equation}
	 (\left<{\boldsymbol{\sigma}}\right> +\left<\tilde{\boldsymbol{\sigma}}\right>)  C^{-1} \boldsymbol{\sigma}
		= \frac{t}{k_B}\boldsymbol{1} \boldsymbol{\sigma} \,.
\end{equation}
As this has to be valid for all  $\boldsymbol{\sigma}$, we conclude that 
\begin{equation}\label{a5}
	\left< \tilde{\boldsymbol{\sigma}} \right> = \frac{t}{k_B} C\boldsymbol{1} -
	\left< {\boldsymbol{\sigma}} \right>  \,,
\end{equation}
Finally, identification of the zeroth-order term leads to
\begin{equation}
	\left<\boldsymbol{\sigma}\right> C^{-1} \left<\boldsymbol{\sigma}\right>= 
	\left<\tilde{\boldsymbol{\sigma}}\right> C^{-1} \left<\tilde{\boldsymbol{\sigma}}\right> \,.
\end{equation}
Using Eq.~\eqref{a5} to eliminate $\left< \tilde{\boldsymbol{\sigma}} \right>$, we then have 
\begin{equation}
	\boldsymbol{1}C\boldsymbol{1} 
	= \frac{2k_B}{t}\boldsymbol{1} \left< {\boldsymbol{\sigma}} \right>\,.
\end{equation}
Finally, noting that $\left<{{\sigma}_1}\right>=J_1 X_1 $ and $\left<{\boldsymbol{\sigma}}\right>=J_2 X_2$, one finds
\begin{align}\label{Call}
	C_{11} &+C_{22}+C_{12}+C_{21} \nonumber \\
		&= \frac{2k_B}{t}
	\left[ X_1^2L_{11}+X_2^2L_{22}+X_1X_2 \left( L_{12}+L_{21} \right) \right] \,.
\end{align}
Since the fluctuation theorem holds for all applied forces, the above result is valid for any $X_1$ and $X_2$.  Setting $X_1$ and $X_2$ separately to zero then leads to 
\begin{equation}
	C_{11} = \frac{2k_B}{t}X_1^2 L_{11} \qquad C_{22} = \frac{2k_B}{t}X_2^2 L_{22} \,.
\end{equation}
Equation~\eqref{CovOns} from the main text follows immediately.

\end{document}